\documentclass[twocolumn]{jpsj2} 
\newcommand{\lsim}
 {\ \raise.35ex\hbox{$<$}\kern-0.75em\lower.5ex\hbox{$\sim$}\ }
\newcommand{\gsim}
 {\ \raise.35ex\hbox{$>$}\kern-0.75em\lower.5ex\hbox{$\sim$}\ }
\def\journal #1#2#3#4{#1 {\bf #2} (#4) #3}
\def\PR{Phys.\ Rev.}
\def\PRB{Phys.\ Rev.\ B}
\def\PRL{Phys.\ Rev.\ Lett.}

\def\JPCM{J.\ Phys.\ Cond.\ Mat.}

\def\JPSJ{J.\ Phys.\ Soc.\ Jpn.}

\def\COMP{Cond.~Mat.~Phys.}
\title{Variational Monte Carlo studies of a $t$-$J$ model on an anisotropic 
triangular lattice}

\author{Tsutomu \textsc{Watanabe}$^{1,2}$\thanks{E-mail: 
h042203d@mbox.nagoya-u.ac.jp}, 
Hisatoshi \textsc{Yokoyama}$^{3}$, Yukio \textsc{Tanaka}$^{1,2}$, 
Jun-ichiro \textsc{Inoue}$^{1}$ and Masao \textsc{Ogata}$^{4}$}

\inst{
$^{1}$Department of Applied Physics, Nagoya University,  Nagoya 464-8603 \\
$^{2}$CREST Japan Science and Technology Corporation (JST) \\
$^{3}$Department of Physics, Tohoku University, Sendai 980-8578 \\
$^{4}$Department of Physics, University of Tokyo, Bunkyo-ku,  Tokyo 113-0033 \\
}

\abst{
With the insulating phase of $\kappa$-$({\rm BEDT}$-${\rm TTF})_2$X 
in mind, we study a $t$-$J$ model on an anisotropic triangular lattice, 
where the hopping integral is $t'$ in one of the three directions, 
using a variational Monte Carlo method. 
By changing the value of $t'/t$, we study the stability of 
superconducting (SC) states with $d$- and $d$+$id$-wave symmetries 
and of an antiferromagnetic (AF) state. 
As $t'/t$ decreases from 1, the stable state immediately switches from
the $d$+$id$ wave to the $d$ wave. 
The AF state is stabilized from the normal spin liquid state for 
$t'/t\lsim 0.7$ at half filling. 
We also take account of Nagaoka ferromagnetism and a phase separation. 
}

\kword{$t$-$J$ model; anisotropic triangular lattice; 
variational Monte Carlo method; $d+id$ wave}

\begin{document}
\maketitle

\section{\label{sec:level1} Introduction}

Recent discoveries of superconductivity in $\kappa$-$({\rm ET})_2$X 
(ET: BEDT-TTF) \cite{Mckenzie} and Na$_x$CoO$_2\cdot y$H$_2$O \cite{Takada} 
are of high interest for some reasons. 
They are strongly correlated electron systems (2$p$-$\pi$ or 3$d$ 
orbitals) with quasi-two-dimensional lattice structures, like 
high-$T_{\rm c}$ cuprates.
However, they differ from the cuprates in that the lattice structure
is essentially triangular, and thus the effect of frustration can 
play a leading role for superconductivity as well as magnetism. 
For such lattices, we expect novel superconducting (SC) features, 
distinguished from the cuprates on the square lattice.
In particular, identification of pairing symmetry is a primary 
subject. 
\par

Recently, we studied the stability of SC states for the $t$-$J$ model 
on the isotropic triangular lattice, using a variational Monte Carlo 
(VMC) method \cite{Watanabe}, which accurately treats the local 
constraint of no double occupation. 
According to it, irrespective of the sign of $t$, a $d_{x^2-y^2}$+$id_{xy}$ 
wave is the most stable among various pairing symmetries near half 
filling, in accordance with mean-field-type approximations \cite{MF}. 
At half filling, the simple $d$ wave is degenerate with the $d$+$id$ wave. 
In contrast, weak-coupling approaches which consider more realistic
band structure of Na$_x$CoO$_2\cdot y$H$_2$O ---a reduced single band 
\cite{single} or multi bands \cite{multi}--- led to different 
symmetries. 
\par

In contrast, $\kappa$-(ET)$_2$X seems to be better described 
using single-band models on an anisotropic triangular lattice, 
where the hopping integral in two of the three directions is $t$ 
and that in the remaining direction $t'$. 
According to the band calculations for $\kappa$-ET salts 
\cite{Komatsu,McKenBand}, the value of $t'/t$ is often near unity 
with $t>0$. 
So far, theoretical studies using the fluctuation exchange (FLEX) 
approximation for the Hubbard model \cite{FLEX} 
have concluded that the pairing symmetry is a simple $d_{x^2-y^2}$ 
wave like the cuprates. 
The results of NMR \cite{NMR} showed that 
the superconductivity is unconventional with nodes, but the pairing 
symmetry is not necessarily $d$ wave. 
\par

In this work, we extend our preceding study \cite{Watanabe} 
to an anisotropic triangular lattice or a $t$-$t'$-$J$-$J'$ model, 
which connects the square lattice ($t'/t=0$) and the isotropic 
triangular lattice ($t'/t=1$). 
The main purpose of this paper is to compare the stability between 
the $d$- and $d$+$id$-wave pairing states, when the values of $t'/t$ 
and carrier density $\delta$ $(1-n)$ are changed. 
At half filling ($\delta =0$), the trial wave functions become insulating 
for the $t$-$J$ model, but the gap parameter $\Delta_k$ for the SC 
state remains finite; thus the favorable form of $\Delta_k$ can be 
determined for this singlet liquid (or RVB) state. 
Actually, $\kappa$-ET salts, whose electron density is often at half 
filling, bring about a superconductor-insulator transition, when their 
effective value of $U/t$ ($U$: onsite repulsion) is increased through 
(chemical) pressure \cite{Mckenzie}. 
Since such a transition is properly described using a Hubbard-type 
model, as shown for the square lattice \cite{YTOT}, we will leave 
it for a coming publication. 
\par

In addition, we also consider the possibility of Nagaoka ferromagnetism 
(FM) and a phase separation.
For the isotropic case with $t<0$, the Nagaoka FM defeats 
the SC states in a wide range of $n$ and $J/|t|$ 
($0\le\delta\lsim 0.96$ and $0\le J/t\lsim 0.6$) \cite{Watanabe,Koretsune}.
It is still controversial whether the phase separation takes place or not 
for very small $J/t$ and $\delta\sim 0$ for the square lattice. 
We thus trace the development of these two states with changing $t'/t$. 
\par

\section{\label{sec:formulation} Formulation}

According to the comparison of experimental results with band 
calculations for the $\kappa$-(ET)$_2$X \cite{Komatsu,McKenBand}, 
the ET molecules in the conducting planes are strongly dimerized.
If each dimer makes a single lattice site, the conducting planes 
can be properly described with a $t$-$J$ model on an anisotropic 
triangular lattice (or an extended square lattice with 
next-nearest-neighbor pairs in only one diagonal direction [1,1]) 
as follows,

\begin{eqnarray}
H=&-&\sum \limits_{\left \langle{i,j} \right \rangle \sigma}t_{ij}
         {P_{\rm G}\left(c_{i\sigma}^\dag c_{j\sigma}+{\rm H.c.}\right)
          P_{\rm G}} \nonumber \\
         &+&\sum \limits_{\left \langle{i,j} \right \rangle}J_{ij}
        {\left({{\bf S}_i\cdot{\bf S}_j-\frac{1}{4}n_i n_j }\right)}, 
\label{eq:model} 
\end{eqnarray}

where $P_{\rm G}=\prod\nolimits_i{(1-n_{i\uparrow}n_{i\downarrow})}$. 
Here, $t_{ij}=t$ and $J_{ij}=J$ for the nearest-neighbor-site pairs 
in two of the three lattice directions, and $t_{ij}=t'$ and $J_{ij}=J'$ 
for the nearest-neighbor-site pairs in the remaining direction, 
and $t_{ij}=J_{ij}=0$ otherwise. 
In view of the relation with the Hubbard model, $J=4t^2/U$, 
we put $J'/J=(t'/t)^2$. 
For $\kappa$-(ET)$_2$X, the value of $t'/t$ is estimated as $0.5$-$1.1$, 
and the electron density is mostly at half filling. 
In this paper, we consentrate on the case of $n\le 1$, because a case 
of $n>1$ can be mapped to a case of $n<1$ with the different signs of 
$t$ and $t'$ through a particle-hole transformation. 
\par

For this model, we perform VMC calculations \cite{Ceperley}, 
which precisely treat the local constraint the $t$-$J$ model imposes. 
As a variational wave function, the Gutzwiller type ($\Psi=P_G\Phi$) 
becomes a good starting point for $t$-$J$-type models \cite{Yokoyama}.
We use $\Psi_{\rm n}=P_{\rm G}\Phi_{\rm F}$ for the normal state 
($\Phi_{\rm F}$: Fermi sea), $\Psi_{\rm SC}=P_{\rm G}\Phi_{\rm BCS}$ 
for the superconducting state ($\Phi_{\rm BCS}$: a BCS function 
with a $\bf k$-dependent gap $\Delta_{\bf k}=\Delta z_{\bf k}$), 
and $\Psi_{\rm AF}=P_{\rm G}\Phi_{\rm AF}$ for the AF state 
($\Phi_{\rm AF}$: a Hartree-Fock state with an AF gap $\Delta$). 
In this paper, we focus on the $d$ wave ($z_{\bf k}=\cos k_x-\cos k_y$) 
and the $d$+$id$ wave 
($z_{\bf k}=\cos k_x+e^{i2\pi/3}\cos (k_x+k_y)+e^{i4\pi/3}\cos k_y$), 
because other symmetries are unstable in the regime of our concern 
for both the isotropic case \cite{Watanabe} and the square lattice \cite{YO}.
In $\Psi_{\rm SC}$ and $\Psi_{\rm AF}$, $\Delta$ is a sole 
variational parameter, which becomes nonzero when $\Psi$ has a long-range 
order, except for the case of $\Psi_{\rm SC}$ at half filling 
($\delta=0$). 
\par

In this study, we collect samples as many as $10^5$-$10^7$, 
which reduce the error in energy to approximately $10^{-4}|t|$.
We use the systems of $N_{\rm s}=8\times 8$-$12\times 12$ with 
periodic-antiperiodic boundary conditions and electron densities 
satisfying the closed shell condition. 
\par

\section{\label{sec:results} Results}

\begin{figure}
\begin{center}
\includegraphics[width=8cm,height=4.5cm]{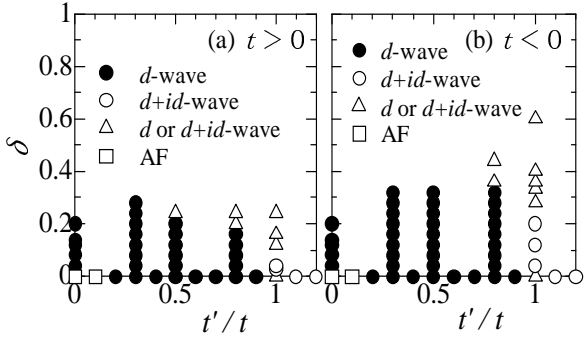}
\end{center}
\caption{
The most stable state among $d$- and $d$+$id$-wave 
and AF states is shown in the $t'/t$-$\delta$ space. 
Here, $J/|t|=0.3$, and the systems are $10 \times 10$ and $12 \times 12$. 
(a) $t>0$ and (b) $t<0$. 
 }
\end{figure}

\begin{figure}
\begin{center}
\includegraphics[width=7.5cm,height=5.5cm]{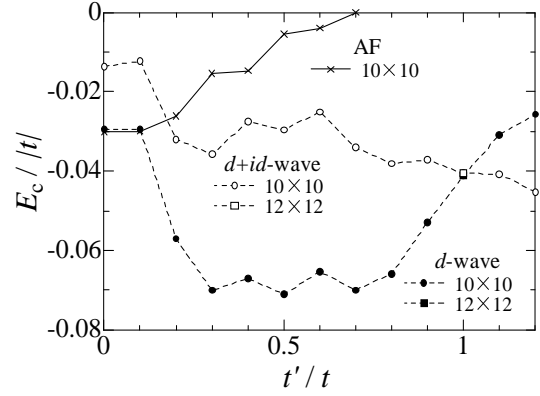}
\end{center}
\caption{
Energy reduction by the $d$- and $d$+$id$-wave spin liquid (or RVB) states 
and the AF state as a function of $t'/t$ at half filling ($\delta =0$). 
The systems are $10 \times 10$ and $12 \times 12$. 
Note that non-smooth behavior of $E_{\rm c}$ chiefly stems from 
the discrete {\bf k}-points of finite systems; when the value of 
$t'/t$ changes, occupied {\bf k}-points (or Fermi surface) discontinuously 
changes for the normal and AF states. 
 }
\end{figure}

Let us start with the phase diagram in the $t'/t$-$\delta$ plain shown 
in Fig.~1, which is constructed by comparing the
energies among the $d$- and $d$+$id$-wave and AF states for $J/|t|=0.3$. 
As studied in \cite{Watanabe}, for the isotropic case ($t'/t=1$), 
the $d$+$id$ wave is the most stable near half filling, but the $d$ wave 
has an indistinguishably close energy for $\delta =0$ and low densities. 
When $t'/t$ is reduced from 1, the stable symmetry rapidly switches 
to the $d$ wave (solid circles), which prevails in a wide range of 
$\delta$ and $t'/t$ including the square lattice, regardless of 
the sign of $t$. 
\par

To consider the half-filled case more in detail, we plot the energy 
difference, $E_{\rm c}=E_{\rm SC}-E_{\rm normal}$, as a function 
of $t'/t$ in Fig.~2. 
Since both $\Psi_{\rm n}$ and $\Psi_{\rm SC}$ are insulating for $\delta =0$, 
$E_{\rm c}\ (<0)$ indicates the energy reduction by forming a singlet 
liquid (or RVB) state \cite{Anderson}. 
When $t'/t$ decreases, $E_{\rm c}$ for the $d$ wave considerably 
decreases, whereas $E_{\rm c}$ for the $d$+$id$ wave tends to increase. 
Thus, the $d$ wave becomes widely stable, especially for $t'/t\sim 0.5$. 
On the other hand, for $t'/t>1$, the $d$+$id$ wave becomes more stable, 
inversely. 
The result of dominant $d$-wave symmetry is basically the same with 
those of the previous FLEX studies \cite{FLEX} for SC states. 
\par

For the square lattice, the AF state $\Psi_{\rm AF}$ is slightly more 
stable than the $d$-wave RVB state $\Psi_{\rm SC}$ at half filling, 
and becomes unstable as soon as holes are doped \cite{YO}. 
Even if $t'/t$ is introduced as in eq.~(1), the cases in
which the AF state is the most stable are limited to half filling 
with $t'/t=0$ and $0.1$, as shown in Fig.~1. 
This is partly due to the fact that the $t$-$J$ model tends to 
underrate the stability of the AF state, because the AF state 
becomes stable in the weaker-coupling region than the $d$-wave does 
\cite{YTOT}. 
However, energy reduction by $\Psi_{\rm AF}$ survives up to 
$t'/t=0.7$ (see Fig.~2). 
Considering, in addition, that for $t'/t=0$ and $0\le\delta\lsim 0.1$, 
a coexisting state of AF and $d$-wave RVB orders is stable \cite{HO}, 
the $d$-wave RVB state at half filling may simultaneously exhibit an 
AF order in some range of $t'/t (>0)$. 
For $t'/t>0.7$, it is probable that a Mott insulator with another 
spin order, e.g.~a $120^\circ$-structure N\'eel order \cite{Capriotti},
appears.
We leave these issues for future studies. 
\par

Next, we discuss Nagaoka FM \cite{Nagaoka}, which is a complete FM 
appearing in general for $J/|t|\sim 0$ and $\delta\sim 0$. 
For the square lattice, its region is fairly narrow, namely,  
$0<\delta\lsim 0.4$ and $J/t\lsim 0.1$ \cite{Ferro}. 
In the isotropic case, the region of FM vanishes for $t>0$, but 
considerably expands to $0\le\delta\lsim 0.96$ and $J/|t|\lsim 0.7$ 
for $t<0$ \cite{Watanabe,Koretsune}.
In Figs.~3, 
we depict phase diagrams for three values of $t'/t$. 
The boundaries are determined by comparing the energies of complete 
FM (spinless fermion) and the $d$ wave (VMC), which is the most stable 
SC state. 
For $t>0$, as $t'/t$ increases, the range of FM rapidly reduces, and 
vanishes for $t'/t=0.5$, whereas for $t<0$, the range of FM rapidly 
expands at first, and becomes almost the same area as the isotropic 
case for $t'/t=0.8$. 
Thus, FM does not seem to appear for $t>0$, whereas for $t<0$ 
the Nagaoka FM is likely to expel the SC state from the realistic 
parameter region even for small $t'/t$. 
\par

\begin{figure}
\begin{center}
\includegraphics[width=6.5cm,height=10cm]{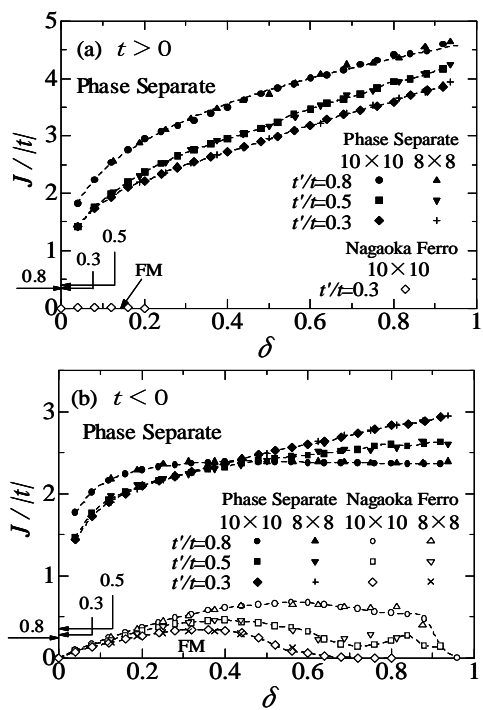}
\end{center}
\vspace{-1cm}
\caption{
Phase diagram in the $\delta$-$J/|t|$ space for the phase separation and 
the complete FM for (a) $t>0$ and (b) $t<0$. 
For $t'/t=0.3$, the range of complete FM with $t>0$ is 
$0\le\delta\lsim 0.2$ and $J/|t|\lsim 0.015$. 
The arrows on the vertical axis indicate the critical values of the 
phase separation estimated by the way (ii). 
 }
\end{figure}

Finally, we consider a phase separation. 
Due to the exchange term, the $t$-$J$-type models necessarily bring about  
a phase separation for large values of $J/|t|$.
Here, we estimate the boundary between the $d$-wave and phase separation 
in two ways.
(i) We suppose that the inhomogeneous phase separates into the two hole 
densities, $\delta=0$ (half filling) and $\delta=1$ (empty), and the 
energy of this state is approximated as $E^{\rm PS}=(1-\delta)E^d$, 
where the $E^d$ is optimized total energy of the $d$ wave at half 
filling.
The boundary is determined by comparing $E^{\rm PS}$ and $E^d(\delta)$. 
Although we use $E^d$ instead of $E^{\rm exact}$ because the latter is 
not known for general values of $t'/t$, the results are connected without 
a marked discrepancy to the cases $t'/t=0$ \cite{YO} and 1 \cite{Watanabe}, 
where $E^{\rm exact}$ is available. 
(ii) We estimate the boundary near half filling, using the condition 
of intrinsic stability, $\partial^2 E/\partial\delta^2>0$, for the $d$ wave. 
Near half filling, this approximation is better. 
\par

The results of (i) and (ii) are summarized in Figure 3.
For $t>0$, the critical value $J_{\rm c}/|t|$ linearly increases as 
$\delta$ increases for large $\delta$, and the slope is slightly 
suppressed, as $t'/t$ decreases.
For $t<0$, as $t'/t$ decreases, the slope of $J_{\rm c}/|t|$ becomes 
steeper.
Regardless of the signs of $t$ and the value of $t'/t$, the estimated 
value of $J_{\rm c}/|t|$ near half filling is small enough to be 
realistic. 
Thus, the possibility of phase separation remains near half filling, 
but this is not the case for intermediate and large values of $\delta$.
\par

\section{\label{sec:conclusion} Conclusion}

We have studied a single-band $t$-$J$ model on an anisotropic 
triangular lattice, based on variational Monte Carlo calculations. 
It is revealed that 
(1) within the superconducting states, the plain $d_{x^2-y^2}$ wave 
is dominant in almost the entire range of $t'/t$ and $\delta$
near half filling.
(2) At half filling, the $d$-wave RVB state is more stable than 
the $d$+$id$ wave for $t'/t<1$, while the relation is opposite 
for $t'/t>1$; the AF state is stabilized from the normal spin 
liquid for $t'/t\lsim 0.7$.
(3) For $t<0$, even when the value of $t'/t$ is small, the Nagaoka 
ferromagnetism occupies a wide range of $\delta$ and $J/|t|$. 
(4) Regardless of the values of $t'/t$, the phase separation may 
arise near half filling, but not for intermediate and large 
values of $\delta$. 
\par

In this work, we have considered a $t$-$J$-type model for the 
insulating phase of $\kappa$-(ET)$_2$X. 
However, we need to tackle a Hubbard-type model to study a
superconductor-insulator transition as well as a superconducting 
phase itself at half filling, like \cite{YTOT} for the square lattice. 
At the final stage in preparing this paper, we became aware of 
a VMC study for a Hubbard model \cite{Trivedi}. 
\par


\begin{acknowledgments}

This work is partly supported by Grant-in-Aids from the Ministry of 
Education, Culture, Sports, Science and Technology, 
by the Supercomputer Center, ISSP, University of Tokyo, 
NAREGI Nanoscience Project, Ministry of Education, Culture, Sports, 
Science and Technology, Japan, which enables us to carry out the calculations 
on the computers at the Research Center for Computational Science, Okazaki 
National Research Institutes, and a Grant-in-Aid for the 21st Century 
COE "Frontiers of Computational Science". 

\end{acknowledgments}



\end{document}